\documentclass[letterpaper,journal]{IEEEtran}
\usepackage{amsmath,amsfonts}
\usepackage{algorithmic}
\usepackage{algorithm}
\usepackage{array}
\usepackage[caption=false,font=normalsize,labelfont=sf,textfont=sf]{subfig}
\usepackage{textcomp}
\usepackage{stfloats}
\usepackage{url}
\usepackage{verbatim}
\usepackage{graphicx}
\usepackage{cite}
\usepackage{hyperref}
\usepackage{xcolor}
\usepackage{float}
\usepackage[acronym]{glossaries}
\usepackage{threeparttable}
\begin{document}
\newacronym{ai}{AI}{Artificial Intelligence}
\newacronym{amr}{AMR}{Autonomous Mobile Robot}
\newacronym{cn}{CN}{Core Network}
\newacronym{cnn}{CNN}{Convolutional Neural Network}
\newacronym{dl}{DL}{Deep Learning}
\newacronym{dt}{DT}{Digital Twin}
\newacronym{gnb}{gNB}{next-generation NodeB}
\newacronym{iiot}{IIoT}{Industrial Internet of Things}
\newacronym{iot}{IoT}{Internet of Things}
\newacronym{lstm}{LSTM}{Long Short-Term Memory}
\newacronym{mba}{MBA}{Microservice-Based Architecture}
\newacronym{ndt}{NDT}{Network Digital Twin}
\newacronym{npn}{NPN}{Non-Public Network}
\newacronym{poc}{PoC}{Proof-of-Concept}
\newacronym{ran}{RAN}{Radio Access Network}
\newacronym{rsrp}{RSRP}{Reference Signal Received Power}
\newacronym{rtt}{RTT}{Round Trip Time}
\newacronym{sa}{SA}{Standalone}
\newacronym{ssb}{SSB}{Synchronization Signal Block}
\newacronym{ue}{UE}{User Equipment}
\newacronym{oai}{OAI}{OpenAirInterface}
\newacronym{kpi}{KPI}{Key Performance Indicator}
\newacronym{ac}{AC}{Application Accuracy}
\newacronym{cfr}{CFR}{Channel Frequency Response}
\newacronym{rrm}{RRM}{Radio Resource Management}
\newacronym{phy}{PHY}{Physical Layer}
\newacronym{5g}{5G}{5th generation}
\newacronym{3gpp}{3GPP}{3rd Generation Partnership Project}
\newacronym{upf}{UPF}{User Plane Function}
\newacronym{los}{LOS}{Line-of-Sight}
\newacronym{nlos}{NLOS}{Non-Line-of-Sight}
\newacronym{tcp}{TCP}{Transmission Control Protocol}
\glsdisablehyper

\title{Digital Twin for Pre-Deployment Validation of AI-Driven Safety-Critical Industrial Edge Control Loops}

\author {Sara~Cavallero, Federico~Tonini, Tony~Chahoud, Giampaolo~Cuozzo, Davide Borsatti, Walter Cerroni, Maurizio~Fodrini, Riccardo~Marini
\thanks{
Copyright (c) 2025 IEEE. Personal use of this material is permitted. However, permission to use this material for any other purposes must be obtained from the IEEE by sending a request to pubs-permissions@ieee.org
\textit{(Corresponding author: sara.cavallero@wilab.cnit.it)}

Sara~Cavallero, Federico~Tonini, Tony~Chahoud, Giampaolo~Cuozzo and Riccardo~Marini are with the National Laboratory of Wireless Communications of CNIT (Wilab, CNIT), Italy. 
(e-mail: sara.cavallero@wilab.cnit.it, federico.tonini@wilab.cnit.it, tony.chahoud@wilab.cnit.it, giampaolo.cuozzo@wilab.cnit.it, riccardo.marini@wilab.cnit.it).
Davide Borsatti and Walter Cerroni are with the DEI, University of Bologna, Italy and WiLab CNIT, Italy (e-mail: davide.borsatti@unibo.it; walter.cerroni@unibo.it).
Maurizio Fodrini is with Fibercop (maurizio.fodrini@fibercop.com).
}}

\markboth{Journal of \LaTeX\ Class Files,~Vol.~14, No.~8, August~2021}%
{Shell \MakeLowercase{\textit{et al.}}: A Sample Article Using IEEEtran.cls for IEEE Journals}


\maketitle

\begin{abstract}

Industrial environments are increasingly characterized by the tight interaction among physical processes, communication infrastructures, and intelligent applications. In this context, Digital Twins (DTs) have emerged as a key technology for system analysis and optimization. However, existing DT solutions typically focus either on industrial processes or communication networks, while lacking an integrated and application-aware perspective. To fill this gap, this paper proposes a modular DT framework for industrial environments that jointly models physical processes, wireless communications, and application logic within a unified architecture. The feasibility of the proposed framework is experimentally validated through a real-world Proof-of-Concept (PoC) implemented in the BI-REX pilot line, involving a 5G-connected Autonomous Mobile Robot (AMR) transporting hazardous liquids and remotely controlled by an AI-driven application. The proposed DT is used to reproduce the behaviour of the real deployment and to investigate the impact of different placements of the AI application, including on-premise, edge, and remote cloud execution scenarios. Experimental results demonstrate a close agreement between DT predictions and PoC measurements in terms of both network-level metrics, such as Reference Signal Received Power (RSRP) and latency, and end-to-end application metrics, including application-level Round-Trip-Time (RTT). Moreover, the analysis shows how inaccuracies of network modeling can critically affect the feasibility of latency-sensitive industrial control loops, highlighting the potential of integrated DTs as tools for the pre-deployment design and validation of next-generation industrial systems.

\end{abstract}


\section{Introduction}

Modern industrial scenarios are evolving toward highly interconnected cyber-physical environments characterized by a continuous exchange of heterogeneous data streams \cite{aceto2019survey}. This phenomenon is enhancing efficiency at the expense of an increased complexity. Indeed, industrial plants typically comprise a large number of heterogeneous devices, including sensors, actuators, programmable logic controllers, industrial robots, autonomous guided vehicles, edge computing nodes, and cloud-based services \cite{cuozzo2025research}. These elements generate data with different temporal granularities, reliability levels, semantic meanings and application requirements. Additionally, industrial processes are intrinsically related, meaning that disturbances occurring in one subsystem can propagate and affect production performance, operational safety, and communication efficiency across the entire facility. Relying on wireless networks for such interconnections is becoming a reality due to their flexibility and scalability, but this further requires careful design, as their performance is influenced by dynamic factors such as interference, mobility, environmental changes, and traffic fluctuations \cite{seferagic2020survey}.

This trend poses significant challenges for system design, optimization, and validation. Experimental evaluations conducted directly on operational industrial infrastructures are often costly, time-consuming, and potentially disruptive to production activities. Furthermore, the large number of interacting parameters makes exhaustive testing under all possible operating conditions practically infeasible.

In this context, \glspl{dt} have emerged as a key enabling technology for the analysis and management of next-generation industrial systems \cite{primo_white_paper_NDT, han2023digital}. A \gls{dt} provides a virtual representation of the physical environment capable of observing, monitoring, and optimizing complex industrial processes while reducing the need for extensive experimentation on the actual plant \cite{soori2023digital}. 

However, most existing \gls{dt} solutions either focus on the process/asset side (e.g., monitoring, predictive maintenance, plant simulation) or on the network/infrastructure side (e.g., performance analysis and optimization, also known as \gls{ndt}~\cite{3gpp28561,conserva2025}), and rarely provide an integrated, application-aware mechanism to jointly model application dynamics (data acquisition, processing, decision, actuation) together with network behavior in a unified, modular architecture. Such an integrated perspective is particularly important for industrial applications relying on closed-loop control mechanisms, whose performance simultaneously depends on physical processes, communication latency, data quality, and application decisions.

To address this gap, this paper proposes and experimentally validates a modular \gls{dt} framework for industrial environments that jointly models industrial processes, wireless communications, and application-level intelligence. The proposed framework combines a Process \gls{dt}, responsible for reproducing the industrial environment and the kinematic behaviour of mobile assets, a Network \gls{dt}, which captures radio propagation and network-level performance, and a Decision Module capable of reproducing the logic and decision-making processes executed by industrial applications.

The proposed framework is validated through a real-world industrial \gls{poc} implemented within the BI-REX pilot line, one of the eight Italian Competence Center for Industry 4.0\footnote{For more information, visit \url{https://bi-rex.it/en/.}}. The considered use case involves an \gls{amr} connected through an \gls{oai}-based \gls{5g} \gls{npn} and transporting hazardous liquids. A remote \gls{ai}-based application logic continuously analyzes telemetry information and predicts future robot movements that may lead to liquid spilling due to abrupt accelerations or braking manoeuvres. Upon detecting a potentially dangerous condition, the application proactively sends a command to the robot to close an onboard clamp and prevent the liquid from being spilled. This use case represents a representative example of an industrial closed-loop control system in which the performance of the overall application critically depends on the interplay between physical processes, communication networks, and application-level intelligence \cite{fragapane2021planning}.

The main contributions of this paper can be summarized as follows:

\begin{itemize}
\item We propose a modular industrial \gls{dt} architecture for industrial environments that jointly models industrial processes, wireless network behaviour, and application logic within a unified framework;

\item We implement the proposed architecture for a real-world industrial use case involving a \gls{5g}-connected \gls{amr}, remote \gls{ai}-based decision making, and proactive safety actions aimed at preventing hazardous liquid spills;

\item We experimentally validate the proposed framework by comparing synthetic predictions generated by the proposed \gls{dt} with measurements collected in the real industrial deployment, considering both end-to-end application metrics, such as \gls{rtt}, and network-level metrics, including \gls{rsrp} and communication latency.

\item We leverage the proposed \gls{dt} to perform a deployment study of the AI-driven application logic of the \gls{poc}, providing quantitative guidelines for the placement of latency-sensitive industrial applications as a function of the accuracy of the network and application modeling.

\end{itemize}

To the best of our knowledge, this is one of the first works providing an experimentally validated industrial \gls{dt} framework capable of jointly modeling physical processes, communication networks, and application intelligence while demonstrating its effectiveness through a real-world industrial deployment.

The remainder of this paper is organized as follows. Section~\ref{sec:related_work} reviews the related literature on industrial \glspl{dt}. Section~\ref{sec:ndt} presents the proposed \gls{dt} framework for industrial environments. Section~\ref{sec:poc} introduces the considered industrial case-study relative to the industrial \gls{poc} developed within the BI-REX pilot line. Section~\ref{sec:dt_implementation} describes the implementation of the proposed \gls{dt} for the considered industrial case-study. Section~\ref{sec:metrics} defines the performance metrics adopted for the \gls{dt} validation. Section~\ref{sec:results} discusses the experimental results obtained by comparing synthetic and real-world measurements. Finally, Section~\ref{sec:conclusions} concludes the paper and outlines future research directions.

\section{Related Work}
\label{sec:related_work}

Recent literature highlights how \glspl{dt} are revolutionizing industry, enabling the digital representation of assets, processes, and networks to optimize operations, predictive maintenance, and resource management \cite{Javaid2023Digital,Xu2023A,Hakiri2024A,hananto2024digital}. Process \glspl{dt} are widely adopted for asset monitoring and predictive maintenance, using real-time data from industrial sensors and data-driven models, to simulate, analyze, and optimize the behavior of industrial assets and plants in domains such as industry and manufacturing \cite{Allahloh2025Integration,Javaid2023Digital,Wahab2024Systematic, Bellavista2021Application-Driven}. In parallel, \glspl{ndt} and \gls{dt}-centric architectures for critical infrastructures and logistics replicate sensors, actuators, and communication technologies to support real-time monitoring, diagnosis, and optimization of networked systems, but often remain focused on maintenance and operations rather than application-layer control \cite{Kherbache2022Network, Hakiri2024A}. Standardization and reference architectures for industrial \glspl{dt} are actively discussed, with surveys and domain reviews highlighting the need for interoperable service‐oriented \gls{dt} stacks and robust communication, computing and security support in industrial domains \cite{Xu2023A,Hakiri2024A}. However, a significant gap remains in the integration between process \glspl{dt} and \glspl{ndt} for the closed-loop optimization of complex industrial systems. Most existing works either focus on process-related aspects (e.g., predictive maintenance and asset optimization) or on the communication infrastructure, and typically adopt relatively monolithic \gls{dt} architectures. In contrast, this paper proposes a modular \gls{dt} framework that tightly integrates process, network, and application-level models within a unified closed-loop architecture. The proposed framework is built upon a fully modular \gls{mba}, where each component of the process and network twins is explicitly decoupled, yet synchronized to jointly support application-level decision making and robotic control. This modular design allows individual components to be independently developed, deployed, updated, scaled, and reused across different industrial applications without redesigning the entire \gls{dt}. Furthermore, unlike most existing studies, the proposed framework is not only presented from an architectural perspective but is also experimentally validated through a real-world industrial \gls{poc}, demonstrating its capability to accurately reproduce both network- and application-level behaviors under realistic operating conditions.

\section{Proposed Digital Twin Framework for Industrial Environments}
\label{sec:ndt}

This section presents the main architectural components of the proposed modular \gls{dt} framework for industrial environments, which jointly models the physical process, the communication infrastructure, and the application logic within a unified modular framework.

\subsection{Architecture Overview}

\begin{figure}
    \centering
    \includegraphics[width=\linewidth]{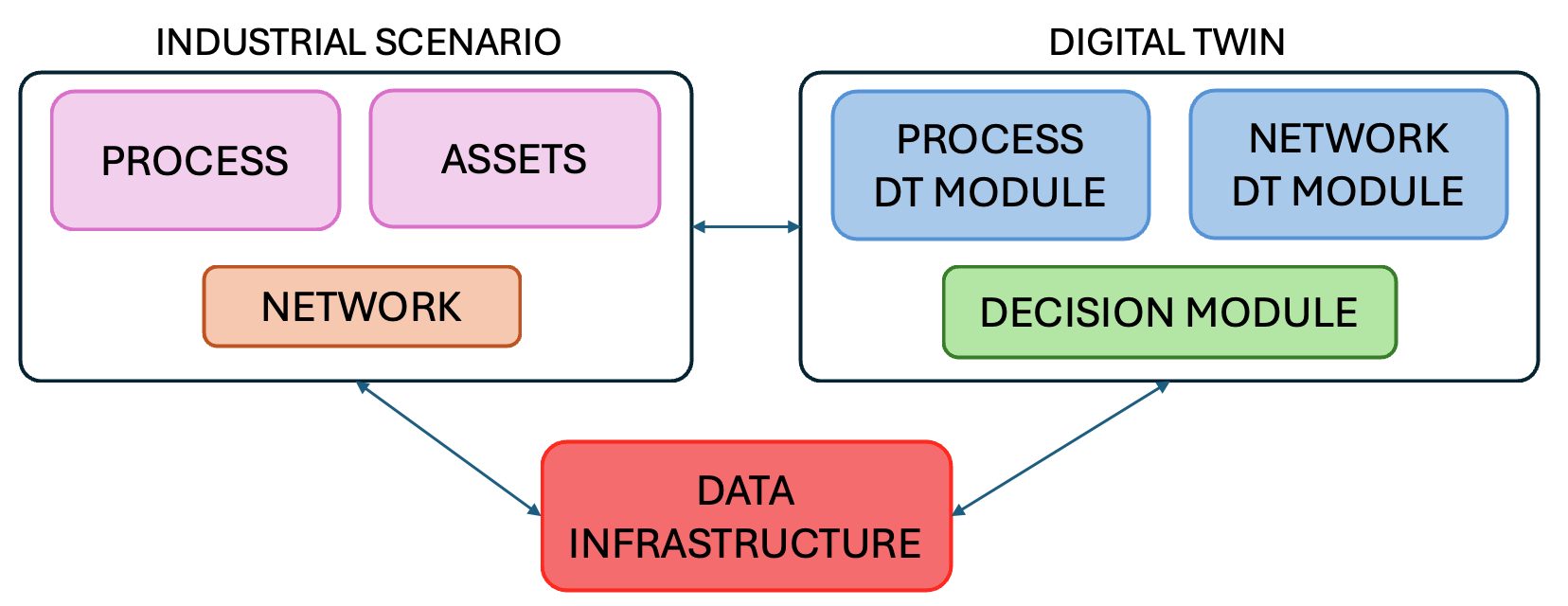}
    \caption{Proposed Digital Twin architecture. The industrial scenario (process, assets, and network) and the DT domain (Process, Network, and Decision Modules) are bridged by a shared data infrastructure. This infrastructure enables a two-way flow: it pulls real-world measurements directly into the \gls{dt} and, in turn, distributes the \gls{dt}'s services, including the synthetic data generated by its modules, to all external domains.} 
    \label{fig:architecture}
\end{figure}

The proposed architecture, depicted in Fig.~\ref{fig:architecture}, shows the interaction between the \emph{industrial scenario} and the \emph{\gls{dt}} domains, mediated by a shared \emph{data infrastructure}. The industrial scenario includes the physical process, the associated assets (e.g., plant geometry, mobile robot kinematics), and the network used to exchange heterogeneous data streams (telemetry, control, etc.). The \gls{dt} domain mirrors this structure through three functional modules: (i) a \emph{Process Digital Twin Module}, or Process \gls{dt}, responsible for reproducing the behavior of the physical process and its assets; (ii) a \emph{Network Digital Twin Module}, or Network \gls{dt}, responsible for reproducing the behaviour of the mobile radio network; 
and (iii) a \emph{Decision Module}, responsible for predictive analytics and proactive network intelligence. The data infrastructure enables a two-way flow: it pulls real-world measurements directly into the \gls{dt} and, in turn, distributes the \gls{dt}'s services, including the synthetic data generated by its modules, to all external domains. The interaction among the three modules enables an iterative design workflow, where information generated by one module is used by the others to evaluate the overall system performance under different deployment configurations and assess the impact of different design choices.

\subsection{Process Digital Twin Module}

The Process \gls{dt} is responsible for modeling the physical domain of the industrial system, including its operational environment, assets, and process dynamics. It provides a digital representation of the system that serves as the foundation for reproducing and analyzing the behavior of the industrial process under different operating conditions.

Building upon this representation, the Process \gls{dt} supports the simulation of process evolution over time, capturing both static and dynamic aspects of the system according to the requirements of the target application. This capability enables the reproduction of realistic operating conditions within a controlled and reproducible digital environment. 

Furthermore, the module can generate synthetic data that complements measurements collected from the physical infrastructure, supporting analysis, validation, and testing activities. The integration of real-world data allows a realistic calibration of such synthetic data generation, ensuring that the simulation outputs remain highly accurate and faithful to reality. 


\subsection{Network Digital Twin Module}

The Network \gls{dt} is responsible for modeling the communication domain of the industrial system and its interaction with the physical process. It reproduces both the communication environment and the underlying network infrastructure, enabling the estimation of communication conditions experienced by connected assets. To this end, the module accounts for propagation effects induced by the operating environment as well as the behavior of the adopted communication technologies and protocols.

Furthermore, the Network \gls{dt} supports the simulation of communication dynamics under different operating conditions and network configurations, enabling the analysis of alternative architectures, protocol settings, and traffic patterns within a controlled and reproducible environment. The resulting network-related information is then integrated with the Process \gls{dt} to provide a unified representation of the industrial system, capturing the mutual influence between process and communication dynamics. This co-simulation framework captures the mutual influence between process and communication dynamics, extending the predictive and risk-mitigation capabilities of the \gls{dt} to the entire industrial ecosystem.

\subsection{Decision Module}

The Decision Module constitutes the predictive intelligence layer of the proposed \gls{dt} framework. It exploits both real-world and synthetic data to train and execute machine-learning models capable of forecasting future system conditions.

The resulting predictions can be exploited by higher-level applications to support proactive decision-making, such as network reconfiguration, resource allocation, trajectory adaptation, or predictive maintenance actions. In this way, the Decision Module complements the descriptive capabilities of the Process and Network \gls{dt} modules with predictive and prescriptive functionalities, enabling a truly intelligent \gls{dt}. Although instantiated in this work through a machine-learning based application, the Decision Module is not tied to a specific \gls{ai} implementation and can generally host application-level logic based on machine learning models, statistical approaches, rule-based algorithms, or other decision-making mechanisms.

\section{Industrial Case-Study}
\label{sec:poc}

In this section, we present a real-world \gls{poc} implemented within the BI-REX pilot line, which serves as the reference application scenario for validating the proposed \gls{dt} framework.

Specifically, we focus on the risks associated with the MIR250 (hereafter referred to simply as MIR), an \gls{amr} \cite{mir250_specs} owned by BI-REX that autonomously navigates the pilot line to transport hazardous liquids (e.g., hydraulic oils) between locations \cite{zhao2023autonomous}. The MIR does not natively include dedicated onboard mechanical equipment to stabilize liquid containers and prevent spills during abrupt maneuvers. Therefore, although the robot is equipped with emergency technologies (e.g., proximity sensors and cameras) that trigger abrupt stops to avoid collisions when a human crosses its path, sudden braking may cause liquid spills, thereby posing a threat to human operators who operate in the same industrial environment.

\subsection{Block Scheme}
To address this issue and improve safety in the BI-REX pilot line, we developed a real-world \gls{poc} based on the interplay between \gls{dl} and 5G.

\begin{figure}[!t]
    \centering
    \includegraphics[width=\linewidth]{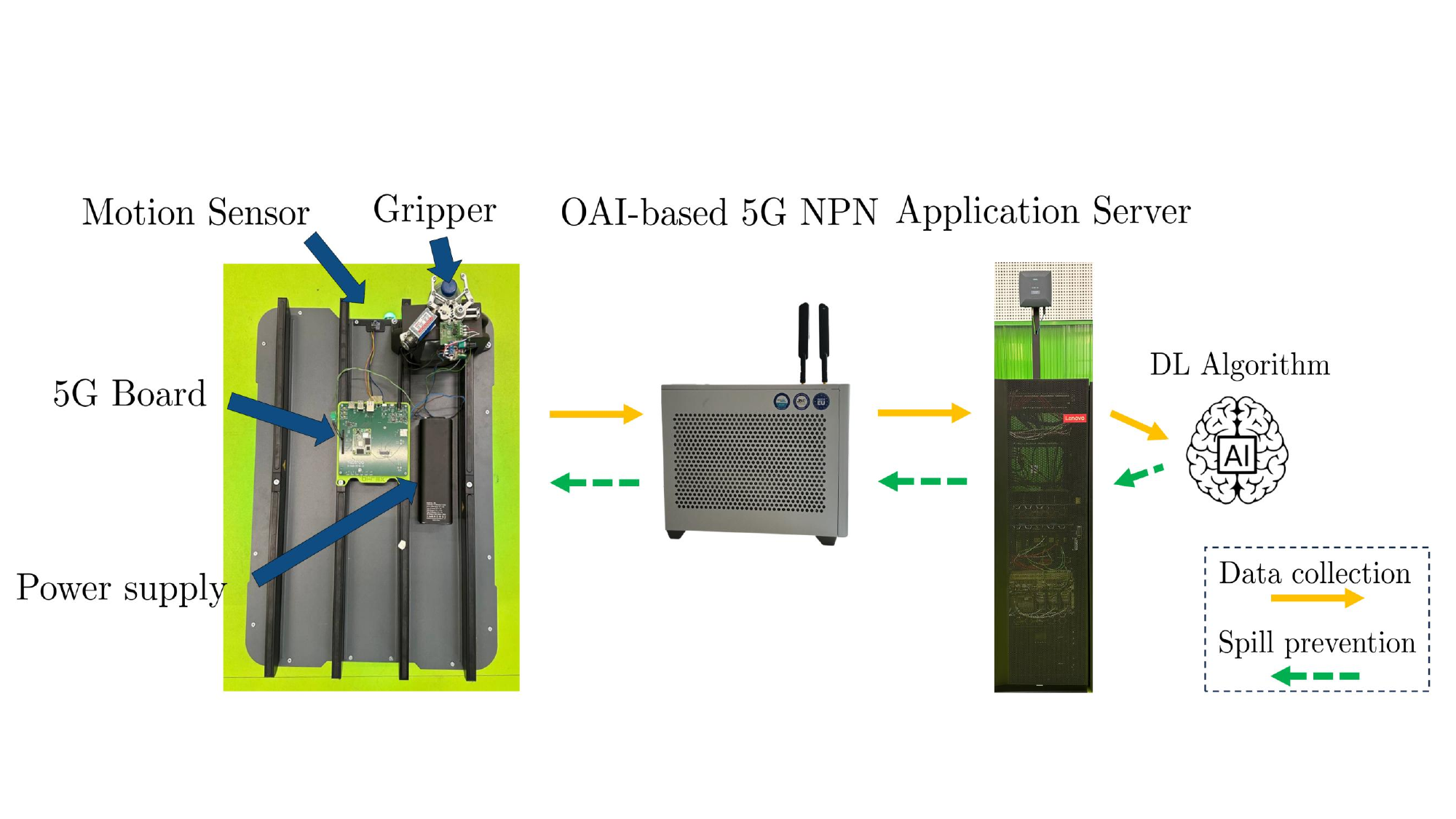}
    \caption{Block diagram of the industrial \gls{poc}.}
    \label{fig:demo_block_scheme}
\end{figure}

Specifically, Fig.~\ref{fig:demo_block_scheme} shows the block diagram of the \gls{poc}. The MIR has been equipped with a motion sensor that collects linear acceleration and angular velocities data with periodicity $T_{\rm S}$. The collected data are transmitted via serial connection to an on-board custom 5G board, which acts as the \gls{ue} in the 5G network and is powered by a power bank as power supply system. 
The board transmits the data with the same periodicity $T_{\rm S}$ to a local application server through an off-the-shelf \gls{oai}-based 5G \gls{sa} \gls{npn}, which is a full-stack and standard-compliant hardware/software platform, consisting of a dedicated \gls{ran} with a \gls{gnb}, and a dedicated \gls{cn}\footnote{More details on the utilized 5G platform can be found \href{https://oaibox.allbesmart.pt/}{here.}}, which have been utilized in the BI-REX pilot line to provide 5G connectivity to the MIR. The local server then hosts an ad hoc \gls{dl} algorithm that performs inference on real-time motion data to predict potential spills. If this is the case, a downlink command is sent to close the custom gripper mounted on the MIR to prevent the dangerous liquid spill.


\subsection{Deep Learning Design}
\label{sec:dl}
As discussed in our earlier theoretical works \cite{amorosa2023end, Longhi2025}, the periodic acquisition of MIR motion data produces an ordered sequence of samples, i.e., a time series. The liquid-spill prediction problem is then modeled as a binary classification task, where a \emph{Fault} event occurs when the MIR path is obstructed and a potential liquid spill can occur. Therefore, the DL model is trained to predict the \emph{Fault} event within an anticipation $A$, where $A$ denotes the number of samples preceding the liquid spill event that can still be labeled as \emph{Fault}. 

Accordingly, each time-series sample is labeled based on its temporal distance from the spill event: the sample corresponding to the spill, together with the $A$ preceding samples, are labeled as \emph{Fault}, while all earlier samples are labeled as \emph{Non-Fault}. 

A compact representation of a labeled time-series is reported in Table~\ref{tab:data_example}, where each row corresponds to motion measurements taken at given time instants, and the temporal distance between two consecutive rows corresponds to the sampling period of $T_{\rm S}$ seconds. In particular, at each timestamp $t \in {1, 2, \dots, t_f}$, where $t_f$ denotes the final timestamp of the time series, i.e., the instant at which the MIR has come to a complete stop, and with respect to the MIR map coordinate frame, the acquired data include i) the linear acceleration vector $\mathbf{a}(t) = (a_x(t), a_y(t))$, where $a_x(t)$ and $a_y(t)$ denote the linear acceleration components along the $x$- and $y$-axes, respectively; ii) the angular velocity component along the $z$-axis, $g_z(t)$; iii) the position vector in Cartesian coordinates $\mathbf{p}(t) = (p_x(t), p_y(t))$; and iv) the orientation angle $\theta(t)$, oriented along the positive $y$-axis of the MIR map frame throughout the predefined trajectory.
\begin{table}[!t]
\centering
\caption{Example of one labeled time-series.}
\setlength{\tabcolsep}{2pt}
\footnotesize
\resizebox{\columnwidth}{!}{%
\begin{tabular}{cccccc}
\hline
$t$ & $\mathbf{a}(t)$ & $g_z(t)$ & $\mathbf{p}(t)$ & $\theta(t)$ & Label \\
\hline
$1$ & $\mathbf{a}(1)$ & $g_z(1)$ & $\mathbf{p}(1)$ & $\theta(1)$ & \textit{Non-Fault} \\
$2$ & $\mathbf{a}(2)$  & $g_z(2)$ & $\mathbf{p}(2)$ & $\theta(2)$ & \textit{Non-Fault} \\
$\vdots$ & $\vdots$ & $\vdots$ & $\vdots$ & $\vdots$ & \raisebox{0.8ex}{\textit{Non-Fault}} \\
$t_f - A$ & $\mathbf{a}(t_f - A)$ & $g_z(t_f - A)$ & $\mathbf{p}(t_f - A)$ & $\theta(t_f - A)$ & \textit{Fault} \\
$t_f - A + 1$ & $\mathbf{a}(t_f - A + 1)$ & $g_z(t_f - A + 1)$ & $\mathbf{p}(t_f - A + 1)$ & $\theta(t_f - A + 1)$ & \textit{Fault} \\
$\vdots$ & $\vdots$ & $\vdots$ & $\vdots$ & $\vdots$ & \raisebox{0.8ex}{\textit{Fault}} \\
$t_f$ & $\mathbf{a}(t_f)$ & $g_z(t_f)$ & $\mathbf{p}(t_f)$ & $\theta(t_f)$ & \textit{Fault} \\
\hline
\end{tabular}%
}
\label{tab:data_example}
\end{table}

The model takes as input a window of $W$ rows from the time series, which is properly preprocessed through feature extraction and normalization steps to preserve short-term temporal dependencies. The window then shifts over the rows (i.e., over time) according to a sliding-window mechanism.\\ 

\begin{figure}[!t]
    \centering
    \includegraphics[width=\linewidth]{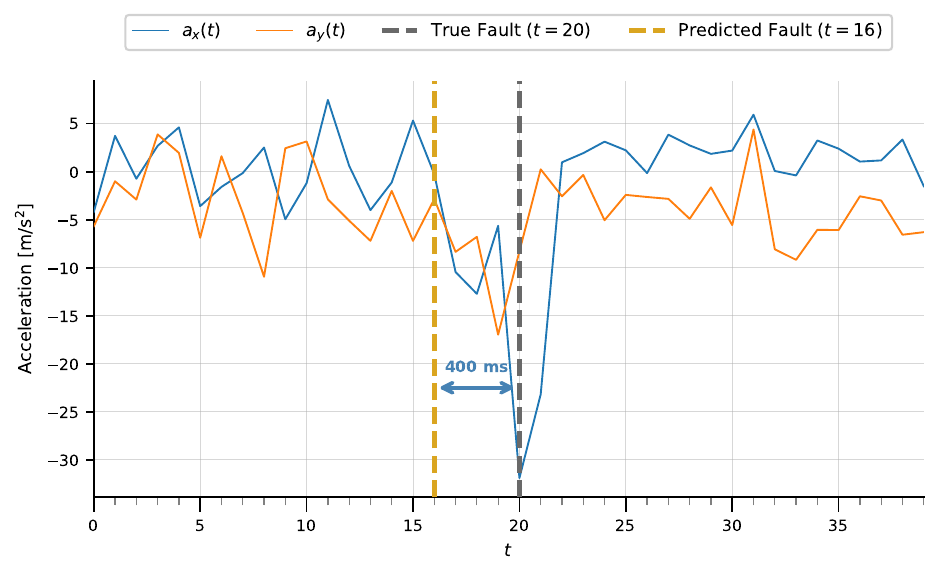}
    \caption{Example of the \emph{Fault} prediction using the trained CNN: the model anticipates the spill event from time series data.}
    \label{fig:cnn_result}
\end{figure}

An example of a \gls{cnn} prediction applied on real-world data is shown in Fig.~\ref{fig:cnn_result}, where $a_x(t)$ and $a_y(t)$ are plotted over the discrete timestamp $t$, consistently with the notation introduced in Table~\ref{tab:data_example}. The yellow vertical line marks the timestamp $t = t_f - A$ at which the \gls{cnn}-based prediction is produced, while the gray vertical line marks the timestamp $t = t_f$ of the actual spill event. In this inference example, with a given sampling period of $T_{\rm S} = 100$~ms, the anticipation time is $400$~ms.

\subsection{Working Condition}
\begin{figure}[t]
    \centering
    \includegraphics[width=\columnwidth]{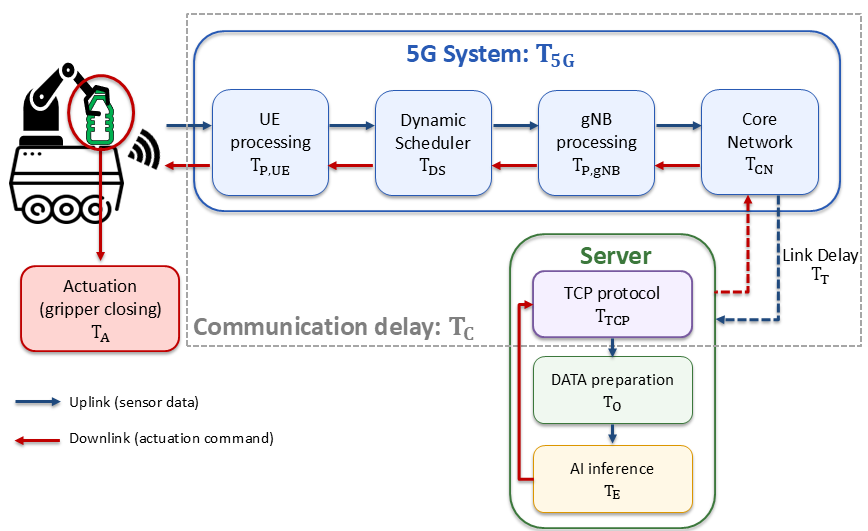}
    \caption{End-to-end communication and processing flow considered in the delay model. Blue arrows denote uplink sensor-data transmission, while red arrows represent downlink actuation commands.}
    \label{fig:latency_scheme}
\end{figure}

Considering a given time series and its first row classified as \emph{Fault} by the \gls{cnn}, the necessary condition for successful operation of the \gls{poc} is given by the following inequality:

\begin{equation}
\label{eq:latency}
T_{\rm I}  \geq T_{\rm C} + T_{\rm O} + T_{\rm E} + T_{\rm A},
\end{equation}

where the different delay contributions and the corresponding communication flow are illustrated in Fig.~\ref{fig:latency_scheme}.

In particular:

\begin{itemize}
  \item $T_{\rm I}$ denotes the application-level delay requirement; in our case, it corresponds to the minimum time available to complete gripper closure without causing a liquid spill;

  \item $T_{\rm C}$ represents the total end-to-end delay introduced within the 5G-based communication system, measured up to the Layer 4 termination points. It is defined as
  $T_{\rm C} = T_{\rm 5G} + T_{\rm TCP} + T_{\rm T}$, and captures the complete closed-loop communication cycle between the \gls{ue} and the network, including uplink transmission of sensor data and downlink delivery of actuation commands from the application server. Specifically,
  $T_{\rm 5G} = T_{\rm P, UE} + T_{\rm P, gNB} + T_{\rm DS} + T_{\rm CN}$, where:
  \begin{itemize}
    \item $T_{\rm P, UE}$ and $T_{\rm P, gNB}$ are the processing delays introduced by the \gls{3gpp} protocol stack at the \gls{ue} and \gls{ran} (\gls{gnb}), respectively, including processing across lower protocol layers;
    
    \item $T_{\rm DS}$ is the scheduling delay introduced by a \gls{3gpp}-compliant dynamic scheduler, where uplink transmissions are scheduled via uplink grants and downlink transmissions are generated in an event-driven manner upon fault detection and scheduled via downlink grants;
    
    
    \item $T_{\rm CN}$ is the \gls{cn} delay, including user-plane processing (e.g., at the \gls{upf}).
  \end{itemize}
  Moreover, $T_{\rm TCP}$ includes the transport-layer (\gls{tcp}) overhead at the endpoints, including packetization, buffering, and reliability mechanisms such as acknowledgments and retransmissions, while $T_{\rm T}$ accounts for the propagation and transmission delay between the \gls{cn} and the application server, which depends on the actual network implementation;

  \item $T_{\rm O}$ quantifies the computational overhead at the application server required to transform raw sensor data into an inference-ready format. This stage ensures data compatibility with the neural network (i.e., the \gls{cnn} in our \gls{poc} case) and involves the reception and organization of incoming data, as well as its conversion into a suitable representation for subsequent processing;

  \item $T_{\rm E}$ denotes the inference time taken by the neural network. It represents the computational effort to generate a final decision, such as classifying the system state as normal operation or fault condition;

  \item $T_{\rm A}$ denotes the actuation delay, i.e., the time required for the gripper to complete the closing action after receiving the control command.
\end{itemize}


\begin{figure}[!t]
    \centering
    \includegraphics[width=\linewidth]{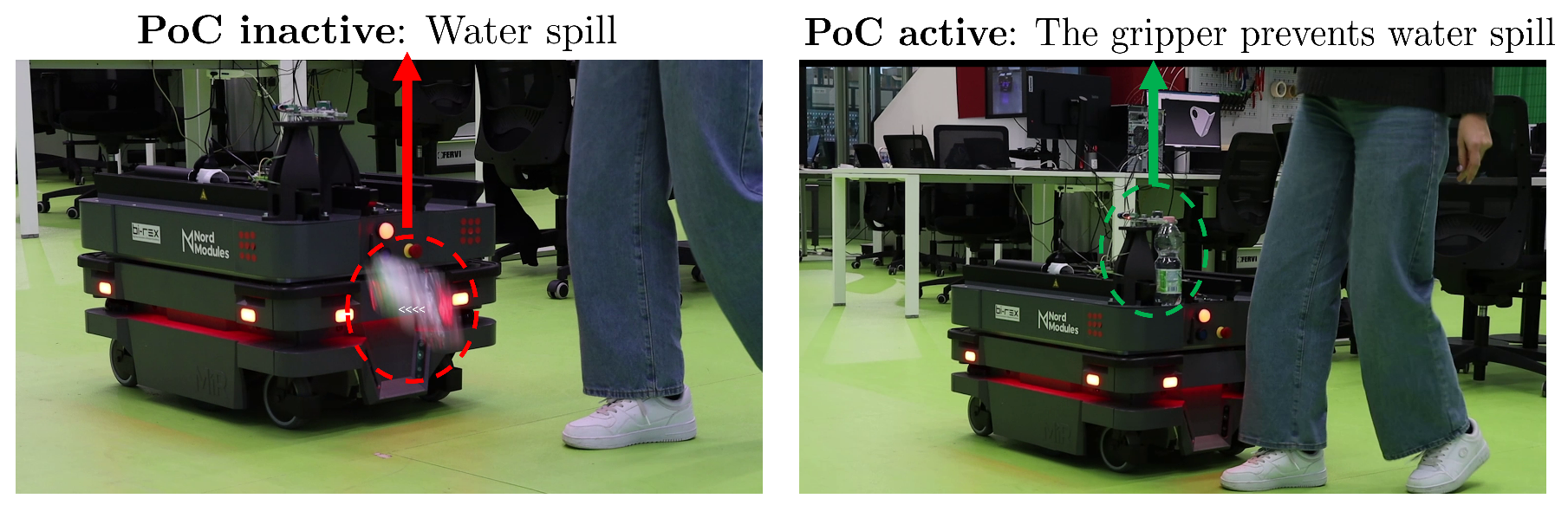}
    \caption{Real-world validation of the industrial PoC. The left image shows the outcome when the \gls{poc} is inactive (i.e., the water bottle falls due to a sudden obstruction), whereas the right image shows that the proposed solution prevents this event by accurately predicting the spill and activating the gripper in advance thanks to the high accuracy of the CNN model and the low latency provided by the OAI-based 5G network.}
    \label{fig:demo}
\end{figure}

To better show how Eq.~\eqref{eq:latency} is used in practice, we consider the toy example shown in Fig.~\ref{fig:cnn_result}. Despite the general formulation in Eq.~\eqref{eq:latency}, the value of $T_{\rm I}$ in our \gls{poc} is empirically estimated at $200$ ms. This is the reason why we set $T_{\rm S}=100$ ms as the reference sampling period. As shown in Fig.~\ref{fig:cnn_result}, the \gls{cnn} predicts a \emph{Fault} $4$ samples in advance, corresponding to an anticipation time of $400$ ms. Since $T_{\rm A}$ was observed to be $50$ ms, and the measured sum of $T_{\rm C}$, $T_{\rm O}$, and $T_{\rm E}$ in our implementation at the BI-REX pilot line is approximately $66$ ms, the \gls{poc} satisfies the timing constraint in Eq.~\eqref{eq:latency} because $T_{\rm I} = 200 > T_{\rm C} + T_{\rm O} + T_{\rm E} + T_{\rm A} = 66 + 50 = 116$ ms. This result is confirmed in Fig.~\ref{fig:demo}, where a photo of the real-world \gls{poc} is depicted. The left image shows the outcome when the \gls{poc} is inactive (i.e., the water bottle falls due to a sudden obstruction), whereas the right image shows that the proposed solution prevents this event by accurately predicting the spill and activating the gripper in real time in advance.



Although the \gls{poc} has been demonstrated within the BI-REX pilot line, with all the components running on a dedicated server, the value of $T_{\rm T}$ depends significantly on the location of the \gls{cnn} (i.e., where it is implemented). It is therefore interesting to understand whether the application can be deployed on-premises, on an edge cloud server, or on a remote cloud server (e.g., one located in a remote data center).
Starting from Eq.~\eqref{eq:latency}, it is possible to derive the following expression:

\begin{equation}
\label{eq:latency-margins}
\begin{aligned}
T_{\rm T}  \leq T_{\rm I} \\
- [(T_{\rm 5G} + T_{\rm TCP} + T_{\rm O} + T_{\rm E}) \cdot (1 + M) + T_{\rm A}],
\end{aligned}
\end{equation}

where $M$ ($\geq0$) represents a margin against the variability of the network and application metrics. This uncertainty margin introduces a robustness against the additional communication and computation uncertainty that can be accommodated while still satisfying the application timing constraint. When the margin is set to $0$, Eq. \ref{eq:latency-margins} sets an upper bound for the \gls{cn}-application server delay, $T_{\rm T}$, which in turn limits the size and distance of the network between the \gls{cn} and the application server. When the uncertainty margin is greater than $0$, the right-hand side of the equation is reduced to account for inaccuracies in estimating the network delays due to variability in data delivery and execution of application instructions/routines on the server platform. Servers located at the edge or remote clouds that satisfy the bound in Eq. \ref{eq:latency-margins} are suitable candidates to host the application workload. While $T_{\rm A}$ and $T_{\rm I}$ depend on the specific application and hardware constraints, $T_{\rm 5G}$, $T_{\rm TCP}$, $T_{\rm O}$, and $T_{\rm E}$ are more difficult to quantify, as they depend on operating conditions such as the environment, location, and traffic load. Conducting extensive real-world experiments to obtain meaningful estimates can be costly and sometimes not feasible; therefore, the use of a \gls{dt} provides a more efficient alternative.

\section{Digital Twin Implementation}
\label{sec:dt_implementation}
\begin{figure}
    \centering
    \includegraphics[width=\linewidth]{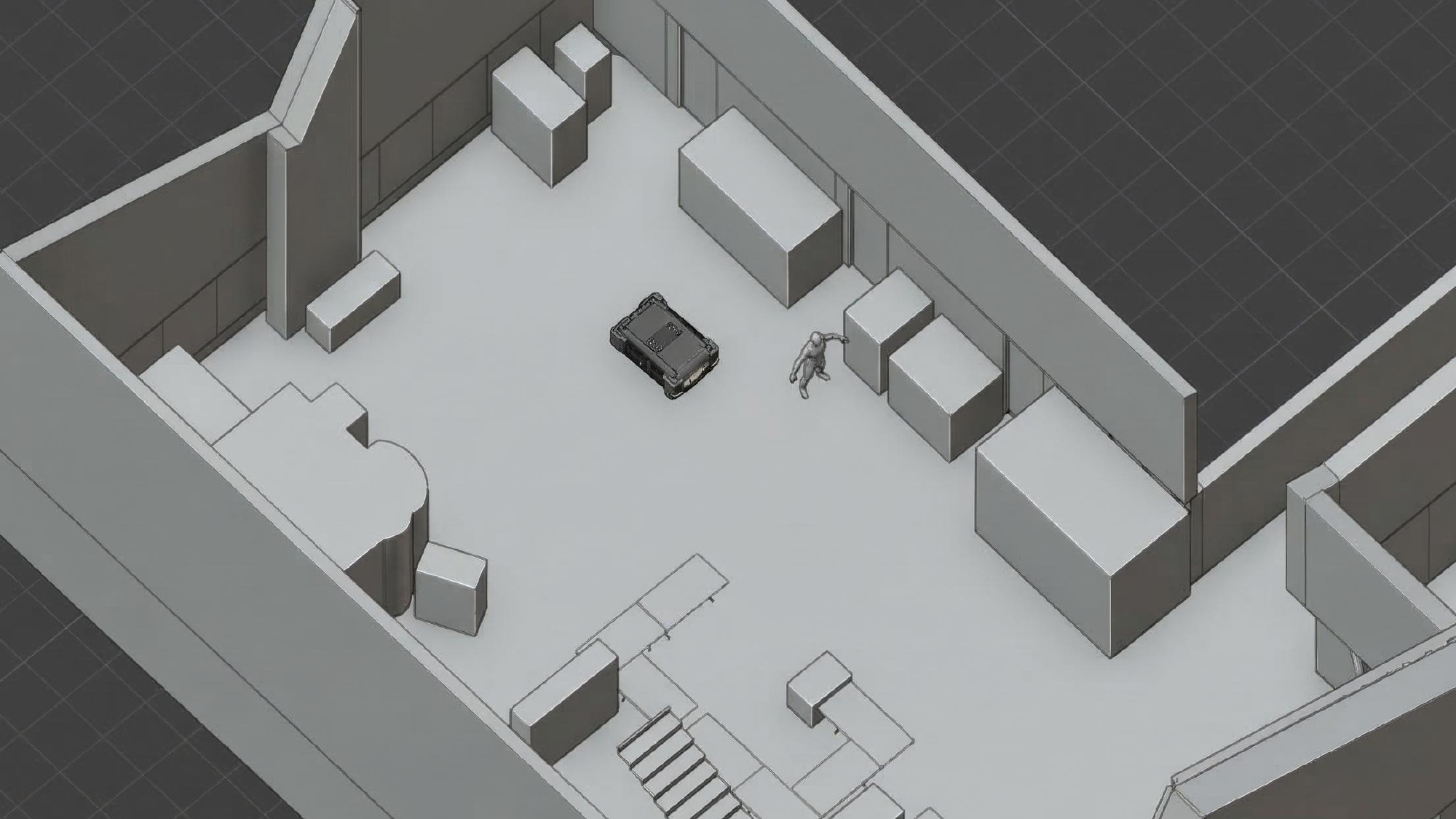}
    \caption{Digital Twin Environment.}
    \label{fig:ndt_3d}
\end{figure}

For the real-world implementation of the \gls{dt} framework described in Section \ref{sec:ndt}, we adopt a \gls{mba} to expose real-time information in a modular and scalable fashion, as already introduced in~\cite{GomezVega2026MSA}. In this way, each functional block is implemented as an independent service that publishes and consumes data through a message broker. In our implementation, RabbitMQ \cite{rabbitMQ} is used as messaging middleware, enabling asynchronous communication and a clean decoupling between data producers (e.g., network interfaces and sensors) and data consumers (e.g., aggregation, inference, and \gls{dt} services).

According to the considered industrial use case, the \gls{dt} framework described in Section ~\ref{sec:ndt} was adapted as follows:

\begin{itemize}
\item The \emph{Process \gls{dt}}, shown in Fig.~\ref{fig:ndt_3d}, has been implemented to provide a complete digital representation of the BI-REX pilot line, obtained through a laser-scanning campaign. The resulting high-fidelity 3D reconstruction serves both as a visualization environment and as a common spatial reference for representing the state of the industrial assets, including the mobile robot. On top of this geometric representation, the Process \gls{dt} implements a mobility model based on the methodology proposed in~\cite{DeNardis2022Mo3}, applied to the digital replica of the MIR. The virtual MIR is configured with the same start and end positions, a reference trajectory spanning approximately 9~m along the pilot line, and a comparable velocity profile as the physical MIR operating in the BI-REX pilot line, ensuring consistency between the digital and physical domains. The model reproduces realistic acceleration and braking dynamics through a goal-seeking heading controller with a first-order speed 
response, including occasional braking irregularities that emulate natural variations in the robot's motion, and triggers abrupt braking maneuvers at the same blockage positions used during the real-world trials.\\
By replaying different operational scenarios within the digital environment, the Process \gls{dt} produces labeled synthetic time series (see Section \ref{sec:dl}) under controlled variations of robot trajectories, speeds, and environmental configurations, which are subsequently exploited to train and evaluate the \gls{ai}-based decision module.

\item The \emph{Network \gls{dt}} simulates the operation of a \gls{3gpp}-compliant \gls{5g} network by exploiting the plant configuration provided by the Process \gls{dt}. First, a 3D ray-tracing engine estimates radio propagation characteristics by accounting for multipath propagation, blockage effects, and \gls{los}/\gls{nlos} transitions induced by the industrial environment geometry, using the well-known Sionna framework presented in~\cite{Hoydis2022Sionna}. The ray-tracing model reproduces the physical deployment by replicating the exact \gls{ue}-\gls{gnb} locations, carrier frequency, and bandwidth, thus generating deterministic, time-varying \gls{rsrp} predictions for each localized position of the MIR along its trajectory. The received signal power is synthesized by combining the contributions of the multipath components computed by the ray-tracing engine, thereby capturing the effects of reflections, diffraction, and shadowing introduced by the industrial environment.

The resulting channel information is then fed into a network-level simulator implementing the main Layer~2 scheduling procedures and message exchanges between the \gls{gnb} and the \gls{ue}. Network communication is managed by a dynamic, \gls{3gpp}-compliant \gls{5g} NR scheduler operating at the \gls{gnb}. Although only one \gls{ue} (the MIR) is considered in this work, the scheduler is designed as a general framework supporting multiple \glspl{ue}; when the available control-plane resources are insufficient, resource allocation is distributed over consecutive scheduling periods until all scheduling requests have been served. The scheduling process alternates between control-plane and data-plane phases, where each uplink or downlink transmission is preceded by the control signaling required for resource allocation. Data-plane resources are allocated according to downlink prioritization, fair resource sharing, and First-In First-Out (FIFO) policies. In the considered industrial scenario, the MIR periodically transmits monitoring data in the uplink, whereas downlink traffic is event-driven and generated only when the decision module detects a fault requiring a notification to be delivered to the robot.


\item The \emph{Decision Module} adopts the same 1D-\gls{cnn} architecture and problem formulation introduced in Section ~\ref{sec:dl}, but is trained and validated entirely on synthetic data generated by the Process \gls{dt} rather than on real-world measurements.
The training dataset is built by simulating a large number of independent virtual \emph{Fault} stops of the virtual MIR at varying trajectory positions $\mathbf{p}(t)$, each characterized by a different realization of the kinematic variables $\mathbf{a}(t)$ and $g_z(t)$ defined in Section ~\ref{sec:dl}, and labeled as \emph{Fault}/\emph{Non-Fault} following the same criterion. Several sliding window sizes $W$ and CNN hyperparameter configurations are explored through a grid search, and the optimal detection threshold is selected via a cost-based optimization that penalizes missed faults more heavily than false alarms \cite{amorosa2023end}.

The trained model is validated on a separate, smaller dataset comprising two distinct, held-out virtual \emph{Fault} stops at different trajectory positions, providing an unbiased assessment of its generalization capability and allowing the \gls{dt} to isolate and quantify the decision delay, including both the preprocessing overhead $T_{\rm O}$ and the inference delay $T_{\rm E}$ under controlled, repeatable conditions.

\end{itemize}

The interaction among the three modules enables the validation of the application logic under different deployment conditions, while supporting what-if analyses through the modification of design parameters and system configurations within the digital environment. This configuration of the \gls{dt} also enables a direct and fair comparison between the predictions generated by the \gls{dt} (i.e., the digital domain) and the measurements collected from the \gls{poc} (i.e., the physical domain). In the following section, we introduce the \glspl{kpi} adopted to quantitatively assess this correspondence.

\section{Performance Metrics}
\label{sec:metrics}




\subsection{Round Trip Time}
We define the \gls{rtt}, $R_{{\rm T},i}$, for the $i$-th \textit{Fault} event as the time elapsed between:
(i) the acquisition time of the first uplink motion sample in the considered time series (see Table \ref{tab:data_example}) that triggers a \emph{Fault} prediction onboard, and
(ii) the reception time, in downlink, of the corresponding gripper closing command onboard.
The \gls{rtt} is computed for each fault event, and the overall performance metric is obtained by averaging over the total number of blockage events considered in the analysis, $N_{\rm E}$, i.e.,

\begin{equation}
 R_{\rm T} = \frac{1}{N_{\rm E}} \sum_{i=1}^{N_{\rm E}} R_{\rm T,i} = \frac{1}{N_{\rm E}} \sum_{i=1}^{N_{\rm E}} (T_{\rm C} + T_{\rm O} + T_{\rm E})_i,
\end{equation}


We emphasize that the three delay terms, $T_{\rm C}$, $T_{\rm O}$, and $T_{\rm E}$, are obtained differently in the \gls{poc} and the \gls{dt}. Specifically, in the real \gls{poc}, these values are measured through experimental campaigns, whereas in the \gls{dt} they are estimated through simulation campaigns based on the network \gls{dt} modeling approach described in Section~\ref{sec:dt_implementation}.

\subsection{Reference Signal Received Power}

We define the \gls{rsrp}, $P_{{\rm S},i}$, for the $i$-th blockage event of the MIR as the received power measured by the \gls{ue} 
within the corresponding event interval. The average \gls{rsrp} over all blockage events $N_{\rm E}$, is then computed as:

\begin{equation}
 P_{\rm S} = \frac{1}{N_{\rm E}} \sum_{i=1}^{N_{\rm E}} P_{{\rm S},i}.
\end{equation}

In the physical \gls{poc}, the instantaneous \gls{rsrp} measurement $P_{{\rm S},i}$ is acquired in real-time through a software module that polls the \gls{ue} using \gls{3gpp} TS 27.007 standardized AT commands.
While in the \gls{dt}, as described in Section \ref{sec:dt_implementation}, the  $P_{{\rm S},i}$ associated with the $i$-th blockage event is computed via the Sionna ray-tracing library \cite{sionna_ref}, mirroring the spatial fading characteristics of the pilot line by replicating the exact \gls{ue}-\gls{gnb} coordinates, frequency, and bandwidth of the physical deployment.

For each localized coordinate associated with the $i$-th blockage event, Sionna calculates the received power by resolving the $K$ multipath components and extracting their respective complex channel gains. Following standard wideband wireless channel modeling principles, the synthesized \gls{rsrp} for the $i$-th \textit{Fault} event is formulated as the incoherent power sum of these paths:

\begin{equation}
P_{{\rm S},i} \; [\text{dBm]}= P_t \;[\text{dBm]} + 10 \log_{10} \left( \sum_{k=1}^{K} \left| H_{k,i} \right|^2 \right),
\end{equation}

where $H_{k,i}$ represents the complex channel gain of the $k$-th propagation path associated with the MIR position during the $i$-th blockage event, and $P_t$ denotes the \gls{gnb} \gls{ssb} transmit power in dBm. This calculation incorporates specular reflections and diffractive effects caused by the specific geometry and material properties of the industrial machinery. 

\section{Results}
\label{sec:results}
Considering the input settings shown in Table \ref{tab:settings}, where the \gls{dt}-related values were determined experimentally, the resulting comparison between \gls{poc} and \gls{dt} is presented in Table \ref{tab:results_comparison}.

\begin{figure}[!t]
    \centering
    \includegraphics[width=\linewidth]{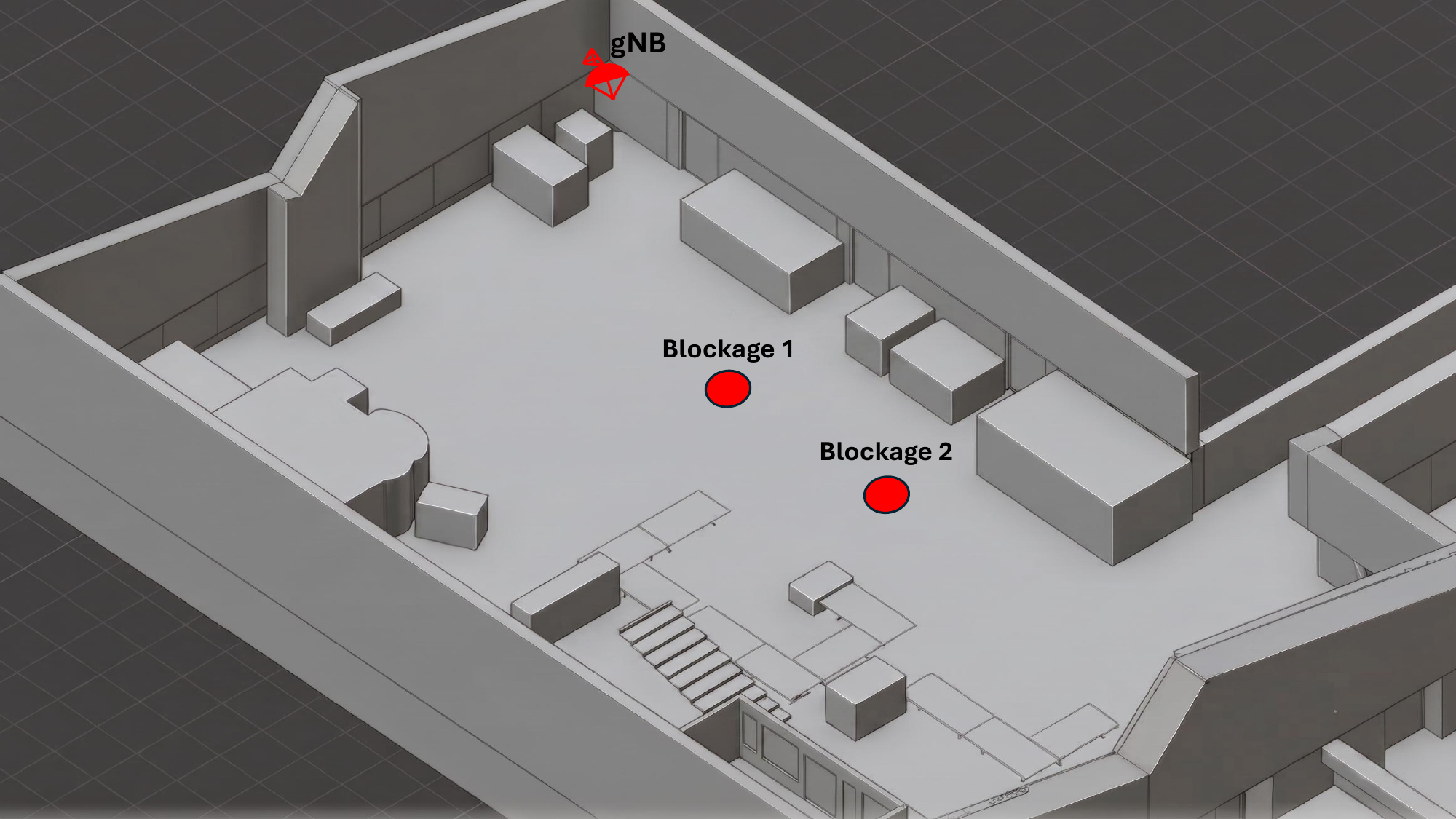}
    \caption{Locations of the two blockage positions used for both \gls{poc} experiments and \gls{dt} evaluations. \emph{Blockage \#1} is placed near the \gls{gnb}, while \emph{Blockage \#2} is farther.}
    \label{fig:stopping_points}
\end{figure}

%
%
\begin{table}[!t]
    \centering
    \caption{Input parameters setting.}
    \label{tab:settings}
    \begin{threeparttable}
    \begin{tabular}{ccc}
        \hline
        Symbol & Description & Value \\
        \hline
        $T_{\rm S}$ & Motion data periodicity & 100 ms \\
        $T_{\rm P,UE}$ & UE processing time & 2 ms \\
        $T_{\rm P,gNB}$ & gNB processing time & 4 ms \\
        $T_{\rm CN}$ & CN delay~\cite{Longhi2025} & 2 ms\\
        $T_{\rm TCP}$ & TCP overhead & 7 ms \\ 
        $T_{\rm T}$ & CN-application server connection delay & 1 ms \\
        $T_{\rm O}$ & Computational overhead at the server side & [25; 45] ms \\
        $T_{\rm E}$ & Inference time & [5; 15] ms\\
        $T_{\rm A}$ & Actuation time & 50 ms \\
        $P_t$ & Transmit power of the SSB signal & 30 dBm \\
        \hline
        \multicolumn{3}{c}{CNN Parameters} \\
        \hline
        $A$ & Anticipation size & 7 \\
        $W$ & Sliding window size & 5 \\
        \hline
    \end{tabular}
    
    \end{threeparttable}
\end{table}
%
%
%
%
\begin{table}[!t]
\centering
\caption{Comparison of the \glspl{kpi} between the \gls{poc} and the \gls{dt} for two blockage events. For the \gls{dt}, results are reported as mean $\pm$ standard deviation over $N_E = 50$, while $N_E = 10$ for the \gls{poc}.}
\label{tab:results_comparison}
\begin{tabular}{|c|c|c|c|c|}
\hline
 & \multicolumn{2}{c|}{PoC} & \multicolumn{2}{c|}{DT} \\
\cline{2-5} 
 & \#1 & \#2 & \#1 & \#2  \\
\hline
$R_{\rm T}$ [ms] & 66.16 & 65.24 & 69.12 $\pm$ 6.13 &  69.21 $\pm$ 6.32\\
$T_{\rm C}$ [ms] & 24.59 & 24.77 & 24.30 $\pm$ 0.80 & 24.25 $\pm$ 0.78\\
$P_{\rm S}$ [dBm] & -77 & -79 & -79 $\pm$ 0.04 & -81 $\pm$ 0.08  \\
\hline
\end{tabular}

\end{table}
Table~\ref{tab:results_comparison} compares the \glspl{kpi} measured in the real-world \gls{poc} with those obtained through the implementation of the proposed \gls{dt} architecture. Two different blockage locations (i.e., Blockage $\#1$ and $\#2$) were considered along the MIR trajectory (see Fig. \ref{fig:stopping_points}), corresponding to a position closer to the \gls{gnb} and a position farther from it, located at approximately 9~m and 7~m from the corner where we placed the \gls{gnb}, respectively. 

Considering the real-world \gls{poc}, $N_{\rm E}=10$ independent experiments were conducted separately for each blockage location by forcing abrupt braking events of the MIR. For each location, the \glspl{kpi} were evaluated as an average over these 10 dedicated runs. Regarding the \gls{dt}, the same two blockage locations were emulated within the virtual \gls{amr} trajectory generated by the mobility model. To this aim, the environment was augmented with a synthetic human obstruction placed at the two predefined positions, where the synthetic \gls{cnn}-based monitoring system consistently detected a blockage event and triggered the corresponding fault prediction. To ensure statistical reliability, the \glspl{kpi} for the \gls{dt} domain were evaluated over $N_{\rm E}=50$ independent simulation runs per blockage position, with the resulting \glspl{kpi} averaged individually for each location. 

Considering the \gls{dt}, the values of $T_{\rm E}$ and $T_{\rm O}$ were randomly sampled within the minimum and maximum bounds observed in the experimental campaign, thus reproducing the variability of the computational overhead and inference stages. Conversely, $T_{\rm C}$ includes both deterministic and implementation-dependent contributions. In particular, $T_{\rm 5G}$ captures the variability introduced by the \gls{3gpp}-compliant scheduler module of the Network \gls{dt} (see Section \ref{sec:dt_implementation}), while $T_{\rm TCP}$ accounts for the transport-layer round-trip overhead induced by the end-to-end communication stack.
Since the proposed Network \gls{dt} focuses on Layer 2 dynamics, higher-layer transport effects are not explicitly modeled within the simulation framework. Therefore, $T_{\rm TCP}$ was independently characterized through dedicated measurements on the real 5G testbed. By comparing \gls{rtt} measurements with and without TCP-related exchange, an average additional delay of approximately 7 ms was estimated. This constant offset was then included in the Network \gls{dt} results to align them with the real \gls{rtt} measurements.

The results reported in Table~\ref{tab:results_comparison} demonstrate a close agreement between the \gls{poc} measurements and the corresponding \gls{dt} predictions. In particular, the average end-to-end communication delay, $T_{\rm C}$, remains nearly constant across the two blockage locations in both the real deployment and the \gls{dt}. This behavior is expected, as the considered scenario involves a single \gls{ue}, resulting in negligible variations in the communication delay due to radio resource scheduling. As shown in Table~\ref{tab:results_comparison}, the standard deviation of $T_{\rm C}$ is below 1~ms, confirming its stability.

Consistently, the average \gls{rtt} predicted by the \gls{dt} differs by less than 4~ms from the corresponding \gls{poc} measurements at both blockage locations. The \gls{dt} also exhibits a standard deviation of approximately 6~ms, mainly due to the randomized values of $T_{\rm E}$ and $T_{\rm O}$, together with the residual uncertainty associated with the complexity of the OAI-based 5G infrastructure. As a result, all experimentally measured \gls{rtt} values fall within the variability predicted by the \gls{dt}, confirming its capability to accurately reproduce the end-to-end delay of the considered application.

The same trend is observed for the \gls{rsrp}, $P_S$. Both the \gls{poc} and the \gls{dt} accurately capture the expected decrease in signal strength from the closer to the farther blockage location, confirming that the framework faithfully replicates the mobile robot's propagation conditions. Although the \gls{dt} exhibits a systematic offset of approximately 2~dBm with respect to the \gls{poc}, the relative variation between the two locations is consistently preserved. As reported in Table~\ref{tab:results_comparison}, the standard deviation of $P_S$ in the \gls{dt} is below 0.1~dB, indicating highly consistent predictions across the simulated blockage events. This discrepancy in absolute values between the \gls{poc} and the \gls{dt} is likely attributable to differences in the \gls{rsrp} estimation procedures adopted in the two cases.

Overall, these results validate the capability of the proposed \gls{dt} to faithfully reproduce the industrial process, 5G network behaviour, and application logic of the real industrial case-study.

\begin{figure}
    \centering
    \includegraphics[width=\linewidth]{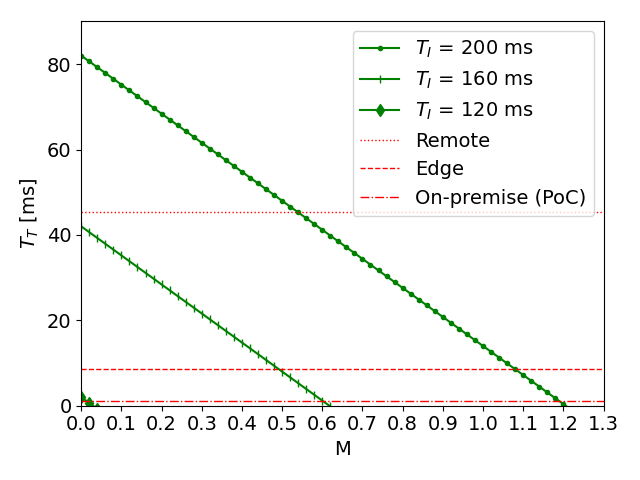}
    \caption{Effects of different margins ($M$) and application requirements ($T_{\rm I}$) on the time budget for $T_{\rm T}$. Reference delay values for different cloud solutions are reported as horizontal lines.}
    \label{fig:Tt}
\end{figure}

Finally, we study the effects of deploying the \gls{dt} services, and more specifically the Decision module, on different servers as a function of different uncertainty margin values (see Fig.~\ref{fig:Tt}).
This analysis allows to understand the maximum margin value that can be assigned to $M$ in Eq. \eqref{eq:latency-margins}, giving an indication of how reliable the estimation on $T_{\rm 5G}$, $T_{\rm TCP}$, $T_{\rm O}$, and $T_{\rm E}$ should be. If $M$ is close to zero, the \gls{dt} estimation of those parameters must be very close to the values in the real world. On the contrary, a large $M$ allows for less accurate \gls{dt} estimations.
Let us consider three cloud options: on-premises, edge cloud, and remote cloud. The on-premises option represents the delay experienced in the actual \gls{poc} deployment within BI-REX pilot line ($1$ ms, see Table \ref{tab:settings}), while the other two correspond to cloud servers located at increasing distances. For the edge and remote cases, the values of the delay correspond to the 90th-percentile measurements over a month for different server locations \cite{cloudping}. These real-world values, reported as horizontal lines on Fig. \ref{fig:Tt}, can be compared with the time budget for $T_{\rm T}$ reported in Eq. \eqref{eq:latency-margins}. More specifically, the figure shows how $T_{\rm T}$ varies with $M$, assuming the equality in \eqref{eq:latency-margins}, indicating the delay margin that is feasible under different application requirements $T_{\rm I}$. When $M = 0$ and $T_{\rm I} = 200$ ms, which is the case of the \gls{poc}, the $T_{\rm T}$ indicates a time budget of about $82$ ms, allowing the workload to be deployed to any cloud server among the ones considered in this study. Increasing $M$ gradually reduces the budget. With $M > 0.55$, the application cannot be deployed on the remote cloud server, as the $T_{\rm T}$ would be lower than the time required to exchange data with that server, violating Eq. \eqref{eq:latency-margins}. When $M > 1.08$, the application workload can be deployed only on-premises. The on-premises option allows for a maximum margin $M = 1.2$.

Decreasing $T_{\rm I}$, corresponding to a tighter industrial requirement, also reduces the overall time budget. When $T_{\rm I} = 160$ ms, the remote cloud servers cannot host the application, while the maximum margin allowed by the on-premises solution is $M = 0.6$.
More stringent application requirements, like the $T_{\rm I} = 120$ ms case, only allow deploying the application on-premises, with almost no uncertainty margin. In this case, edge and remote cloud servers are not suitable to host the application.


\section{Conclusions}
\label{sec:conclusions}

This paper presented a modular \gls{dt} framework for industrial environments, capable of jointly modelling industrial processes, wireless communications, and application logic within a unified architecture. The proposed framework integrates a Process \gls{dt}, a Network \gls{dt}, and an \gls{ai}-oriented Decision Module, enabling the end-to-end representation of complex industrial cyber-physical systems.

To demonstrate the practical applicability of the proposed approach, the framework was implemented and experimentally validated through a real-world \gls{poc} deployed within the BI-REX pilot line and according to an \gls{mba}. The considered use case involved a \gls{5g}-connected MIR transporting hazardous liquids and remotely controlled by an \gls{ai}-driven application capable of predicting potentially dangerous manoeuvres and proactively triggering in advance a safety closure of the onboard gripper.

The experimental results showed a close agreement between the synthetic predictions generated by the \gls{dt} and the measurements collected during the field trials. In particular, the average prediction error was equal to 5\% for the \gls{rtt}, 2\% for the end-to-end communication latency, and a maximum absolute deviation of 2~dB for the \gls{rsrp}, demonstrating the capability of the proposed framework to accurately reproduce both network-level and application-level behaviours.

Furthermore, the proposed \gls{dt} was leveraged as a design-space exploration tool to investigate the impact of different placements of the \gls{ai} application logic of the \gls{poc}, namely on-premises, edge cloud, and remote cloud servers. The analysis showed that the considered industrial control loop remains feasible even under significant uncertainty in the estimated communication and application delays. In particular, the on-premises, edge, and remote cloud deployments can accommodate uncertainty margins of up to $1.2$, $1.08$, and $0.55$, respectively, before violating the application timing requirements. These results demonstrate the potential of the proposed framework to support deployment decisions for latency-sensitive industrial applications by evaluating not only nominal performance, but also their resilience to uncertainty.

Overall, the presented results indicate that integrated industrial \gls{dt} can serve not only as accurate replicas of real-world systems but also as effective tools for what-if analyses, proactive system design, and deployment optimization. It is worth noting that the deployment analysis presented in this work focuses on latency-related aspects and their impact on industrial control-loop feasibility. Other deployment considerations, such as scalability, operational costs, resource availability, and resiliency, are outside the scope of the present study. Future research directions will focus on extending the proposed framework with online synchronization mechanisms, closed-loop optimization capabilities, and \gls{ai}-driven autonomous orchestration functions to enable self-adaptive industrial cyber-physical systems.

\section{Acknowledgments}
This work was partially supported by the European Union under the Italian National Recovery and Resilience Plan (NRRP) of NextGenerationEU, partnership on “Telecommunications of the Future” (PE00000001 - program “RESTART”) and PON-SUNS - Secure Ubiquitous Networks and Services MI\_1301 Avviso D.D. 307 2025 - CUP B89H26000100007 - Azione: 1.1.2.


\bibliographystyle{IEEEtran}
\bibliography{IEEEabrv,bibliography}

\begin{IEEEbiography}[{\includegraphics[width=1in,height=1.25in,clip,keepaspectratio]{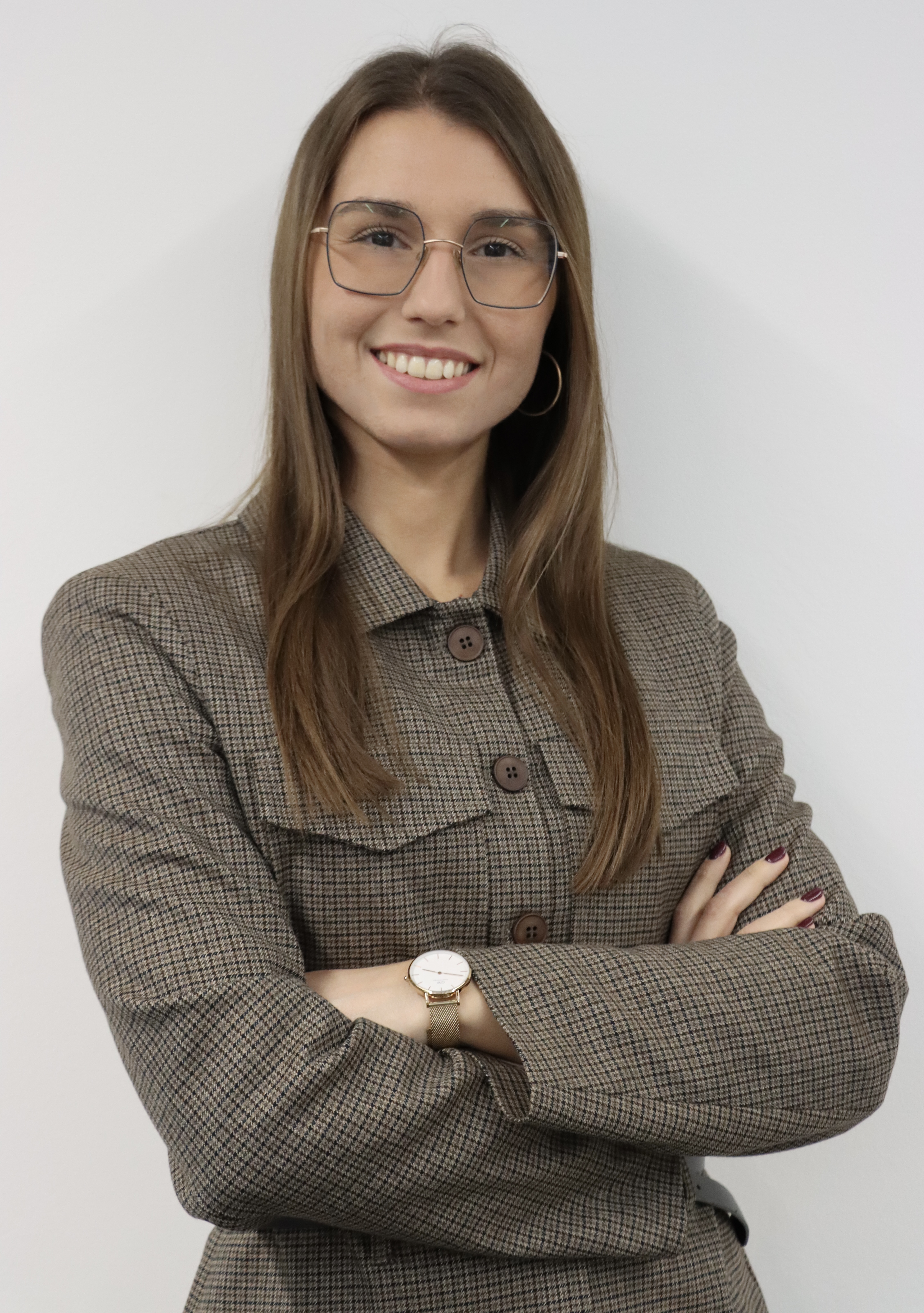}}]{SARA CAVALLERO } pursued the Ph.D. in Electronics, Telecommunications, and Information Technologies Engineering (ETIT) at the University of Bologna in 2025, where her research focuses on the design and modeling of MAC protocols for Industrial Internet of Things applications operating at THz frequencies, the development of scheduling optimization algorithms for 5G NR networks to support URLLC, and work on Network Digital Twin technologies. In recognition of her research excellence, she received the Francesco Carassa Award in 2024. Previously, in 2022, she was awarded the “Mela d’Oro” in the New Graduate category by the Marisa Bellisario Foundation for the best thesis in Telecommunications Engineering.
\end{IEEEbiography}

\begin{IEEEbiography}[{\includegraphics[width=1in,height=1.25in,clip,keepaspectratio]{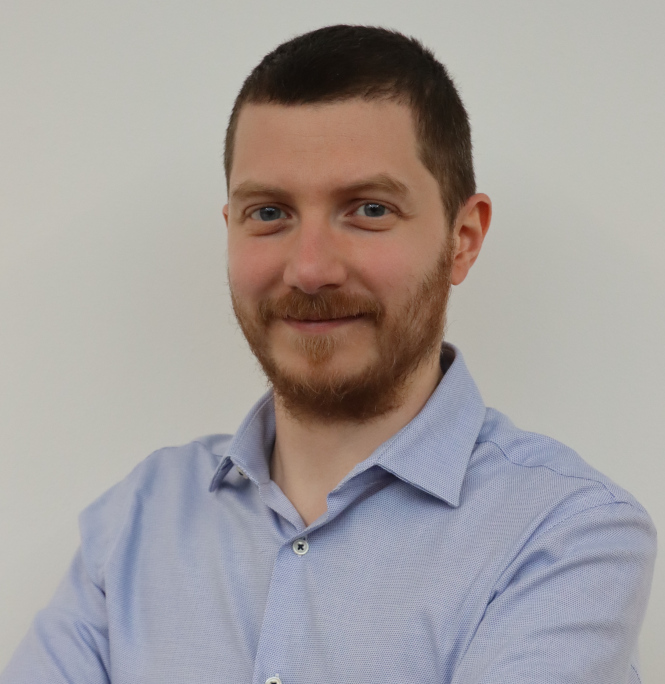}}]{FEDERICO TONINI }(Member, IEEE) received the Ph.D. degree in Electronics, Telecommunications, and Information Technology Engineering from the  University of Bologna, Italy, in 2019. He is a Researcher with the National Inter-University Consortium for Telecommunications (CNIT) at the Wireless Laboratory (WiLab), Bologna, Italy. His research interests include mobile network orchestration, transport network design, and resource optimization techniques.
\end{IEEEbiography}

\begin{IEEEbiography}[{\includegraphics[width=1in,height=1.25in,clip,keepaspectratio]{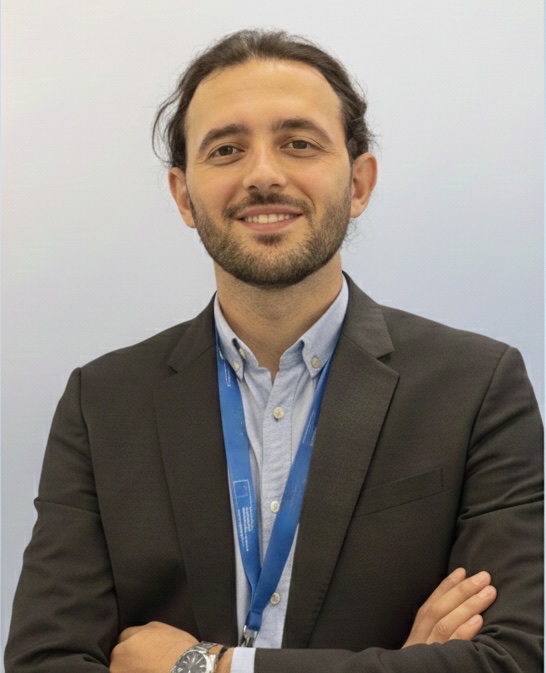}}]{TONY CHAHOUD } (Member, IEEE) is a Researcher at WiLab, the National Laboratory of Wireless Communications of CNIT, Italy. He received the Engineering Degree in Computer and Communications Engineering from the American University of Science and Technology, Beirut, Lebanon, in 2020. In 2024, he received the M.S. degree in Artificial Intelligence from the University of Bologna, Italy. His research interests broadly span machine learning and its applications to wireless communications and Industrial IoT.
\end{IEEEbiography}

\begin{IEEEbiography}[{\includegraphics[width=1in,height=1.25in,clip,keepaspectratio]{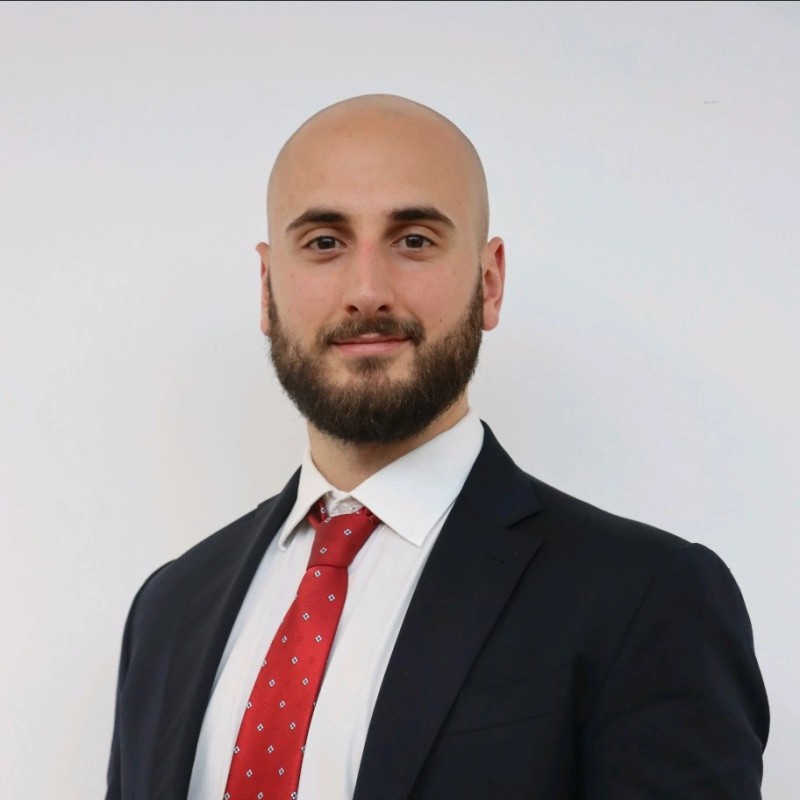}}]{GIAMPAOLO CUOZZO } received the B.S. degree (Hons.) in electronics and telecommunications engineering from the University of Bologna in 2017, the M.S. degree (Hons.) in telecommunications engineering in 2019, and the Ph.D. degree in electronics, telecommunications, and information technologies engineering (ET-IT) in 2022. He is currently the Head of Research with the National Laboratory of Wireless Communications (WiLab), CNIT (the National, Inter-University Consortium for Telecommunications). His research interests focus on the design, development, and evaluation of agentic and deep learning algorithms for beyond-5G networks, with particular emphasis on multiple access techniques, scheduling procedures, and their validation in Industrial Internet of Things (IIoT) environments. He is the author of one book, one invention, and more than twenty articles.
\end{IEEEbiography}

\begin{IEEEbiography}[{\includegraphics[width=1in,height=1.25in,clip,keepaspectratio]{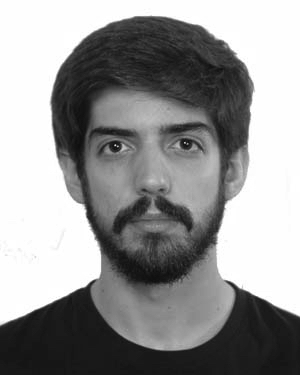}}]{DAVIDE BORSATTI } (Member, IEEE) received his B.S., M.S., and Ph.D. in Electronics, Telecommunications, and Information Technologies Engineering from the University of Bologna, Italy, in 2016, 2018, and 2022, respectively. He is currently an Assistant Professor in the Department of Electrical, Electronic, and Information Engineering "Guglielmo Marconi" at the University of Bologna. His research interests include NFV, SDN, intent-based networking, MEC, and 5G network slicing.
\end{IEEEbiography}

\begin{IEEEbiography}[{\includegraphics[width=1in,height=1.25in,clip,keepaspectratio]{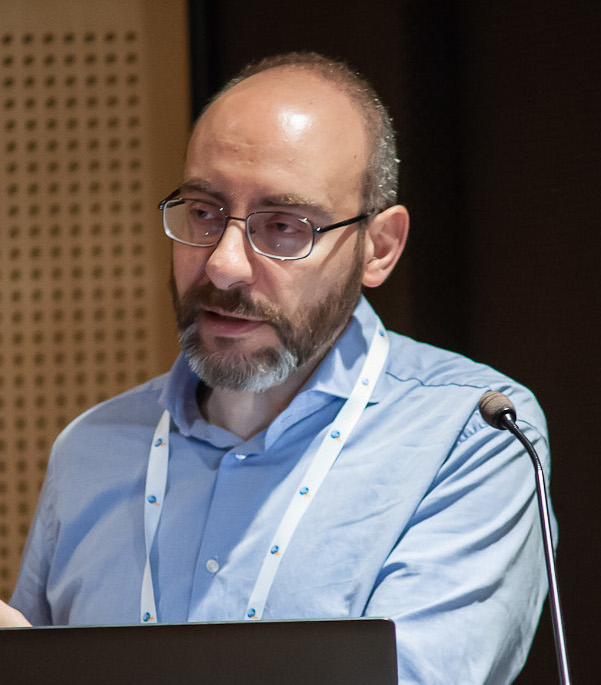}}]{WALTER CERRONI } (Senior Member, IEEE) is an Associate Professor of communication networks with the Department of Electrical, Electronic, and Information Engineering ``G. Marconi'' of the University of Bologna, Italy. He coauthored more than 150 articles published in the most renowned international journals, magazines, and conference proceedings. His recent research interests focused on multiple aspects of control, management and orchestration of communication network infrastructures, including software-defined networking, network function virtualization, multi-access edge computing, fog computing, service function chaining, intent-based networking systems. He serves/served as Series Editor for the IEEE Communications Magazine, Associate Editor for the IEEE Communications Letters, and Technical Program Co-Chair for IEEE-sponsored international workshops and conferences.
\end{IEEEbiography}

\begin{IEEEbiography}[{\includegraphics[width=1in,height=1.25in,clip,keepaspectratio]{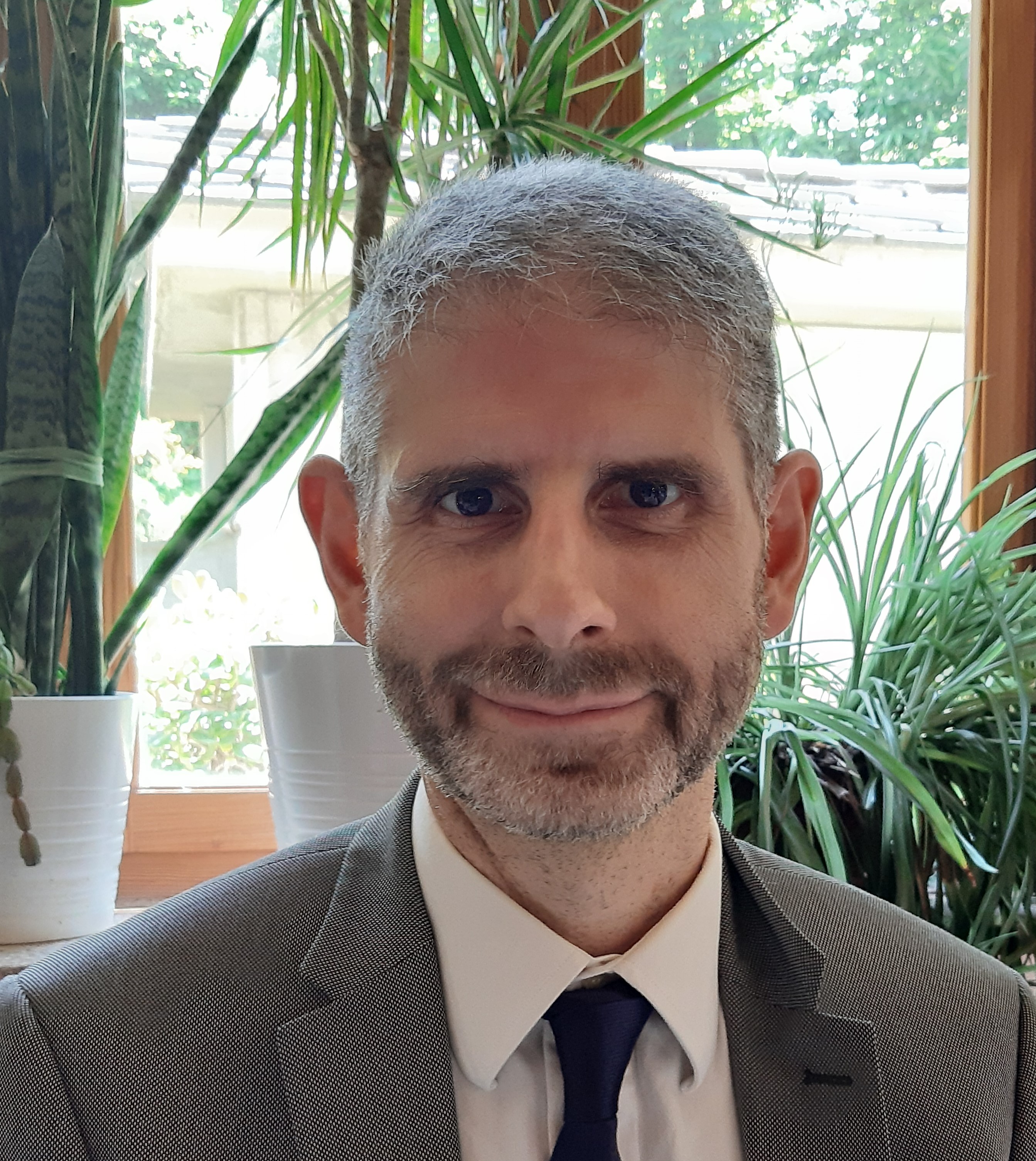}}]{MAURIZIO FODRINI } graduated in Telecommunications Engineering, he joined Telecom Italia in 2001. Since July 2024, he has been part of FiberCop, where he works within the Technology Plans and Innovation Programs department. In this role, he is involved in innovation initiatives, with a particular focus on collaborations with universities, national and European research projects, network automation, and the evolution towards autonomous networks. He also contributes to 3GPP standardization activities, particularly in the context of 6G.
Throughout his career, he has played a key role in the architectural evolution of mobile network management and orchestration systems, the development of simulation platforms for evaluating radio system performance, and the design of multimodal technologies for advanced human-machine interaction. He has also been involved in the scouting, benchmarking, and trialing of emerging radio access technologies and mobile network architectures, contributing to the definition of the Group's Technology Plan.
Within 3GPP, he has contributed to the standardization of radio interfaces for 3G, 4G, and 5G systems. He also served as Network Leader of the "TIM Open Knowledge" community, promoting knowledge sharing and collaboration across the organization.
Over the years, he has contributed to the development of several patents, the preparation of RFI and RFQ documents, and the execution of activities within European research projects. He has participated in international conferences and delivered training programs for the Group's subsidiaries in South America.
\end{IEEEbiography}

\begin{IEEEbiography}[{\includegraphics[width=1in,height=1.25in,clip,keepaspectratio]{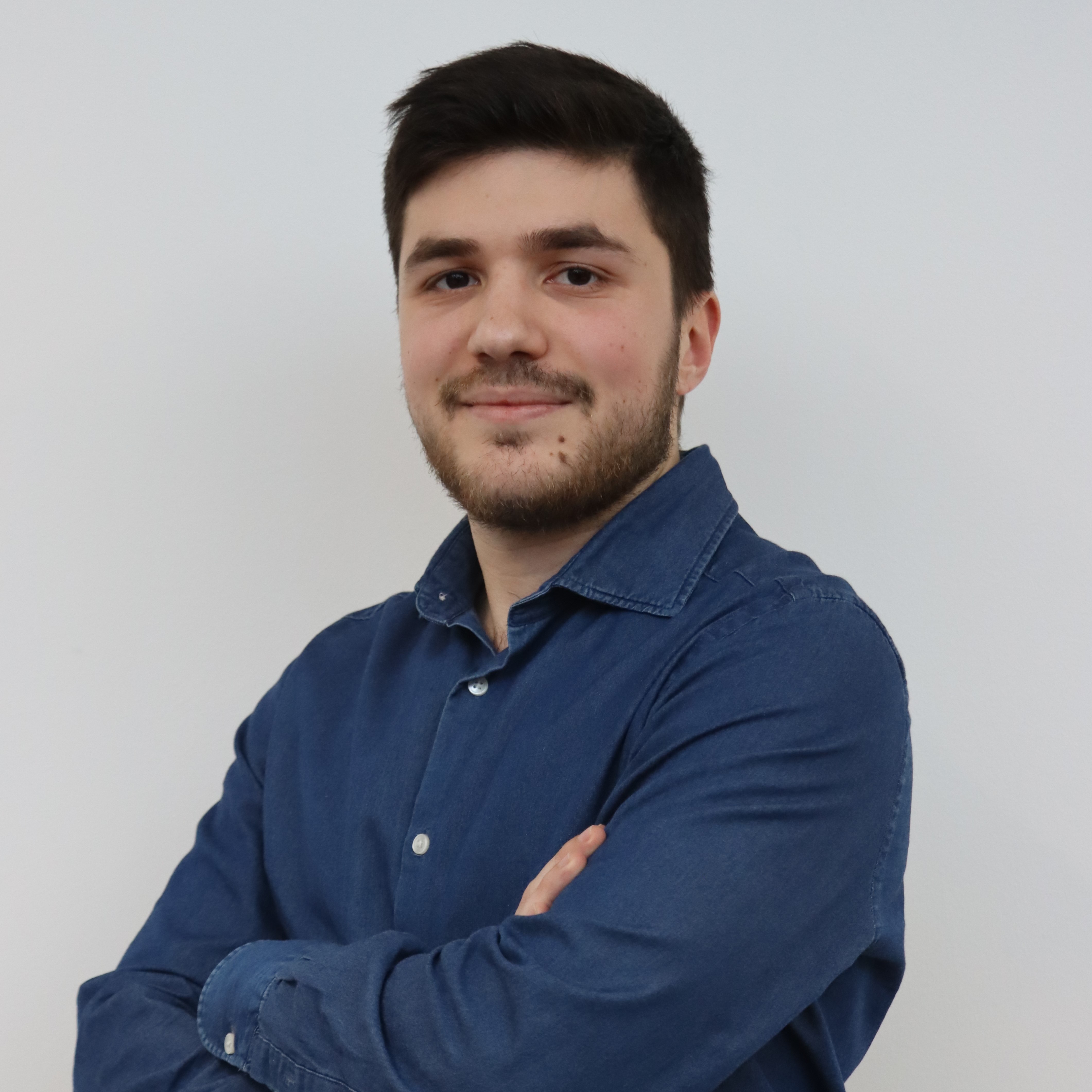}}]{RICCARDO MARINI } is Head of Research at WiLab, the National Laboratory of Wireless Communications of CNIT, where he serves as Sector Leader for Research and Knowledge Transfer. He received his M.Sc. degree in Telecommunications Engineering in 2019 and his Ph.D. in Electronics, Telecommunications and Information Technology in 2023 from the University of Bologna. His research interests include Network Digital Twins, Internet of Things, non-terrestrial networks, and drone communications. He coordinates research activities with industrial partners, focusing on the design and validation of next-generation network architectures, protocols, and proof-of-concepts.
\end{IEEEbiography}

\end{document}